\documentclass{article}
\usepackage[margin=1in]{geometry}
\usepackage[super]{natbib}
\usepackage{cite}
\usepackage[pdftex]{graphicx}
\DeclareGraphicsExtensions{.pdf,.png}
\usepackage{array}
\usepackage{fixltx2e}
\usepackage{stfloats}
\usepackage{setspace}
\usepackage{amsmath}
\usepackage{url}

\begin{document}
\pagenumbering{gobble}

\begin{centering}
\clearpage\thispagestyle{empty}
\textbf{\Large{Predicting Instabilities in Transient Landforms and Interconnected Ecosystems}} 
\vspace{2cm}

Taylor Smith$^{1}$, Andreas Morr$^{2,3}$, Bodo Bookhagen$^{1}$, Niklas Boers$^{2,3,4}$\\
$^{1}$Institute of Geosciences, Universit\"{a}t Potsdam, Germany\\
$^{2}$ Earth System Modelling, School of Engineering \& Design, Technical University of Munich, Germany  \\
$^{3}$Potsdam Institute for Climate Impact Research, Germany\\
$^{4}$Department of Mathematics and Global Systems Institute, University of Exeter, UK\\

\end{centering}

\vspace{16cm}

\noindent
Corresponding author: \\
Taylor Smith \\
Email: tasmith@uni-potsdam.de

\clearpage
\doublespacing

\begin{abstract}
Many parts of the Earth system are thought to have multiple stable equilibrium states, with the potential for rapid and sometimes catastrophic shifts between them. The most common frameworks for analyzing stability changes, however, require stationary (trend- and seasonality-free) data, which necessitates error-prone data pre-processing. Here we propose a novel method of quantifying system stability based on eigenvalue tracking and Floquet Multipliers, which can be applied directly to diverse data without first removing trend and seasonality, and is robust to changing noise levels, as can be caused by merging signals from different sensors. We first demonstrate this approach with synthetic data and further show how glacier surge onset can be predicted from observed surface velocity time series. We then show that our method can be extended to analyze spatio-temporal data and illustrate this flexibility with remotely sensed Amazon rainforest vegetation productivity, highlighting the spatial patterns of whole-ecosystem destabilization. Our work applies critical slowing down theory to glacier dynamics for the first time, and provides a novel and flexible method to quantify the stability or resilience of a wide range of spatiotemporal systems, including climate subsystems, ecosystems, and transient landforms. 
\end{abstract}

\clearpage
\newpage

\section*{Introduction}

The study of abrupt transitions has been widely applied to different parts of the Earth system in recent years \citep{Lenton2008,boers2022_erl,lenton2024}, including global vegetation systems \citep{smith2022,forzieri2022}, ice sheets \citep{boers2021}, and ocean circulation systems \citep{boers2021b,shin2025}. A suite of methods has been proposed to provide early warnings of such transitions, with the most prominent ones relying on the concept of `critical slowing down' (CSD) based on slowing dynamics around critical transition points \citep{Carpenter2006,Dakos2008,Scheffer2009,boers2022_erl}. Most studies have relied on either a single time series of a key state variable (e.g., paleoclimate proxies \citep{Dakos2008,Boers2018,boers2022_erl}) or sets of time series representing changes of a key variable (e.g., vegetation productivity \citep{smith2022,forzieri2022}) assessed individually. Early warning signals are often measured using the most common CSD indicators -- lag-one autocorrelation (AC1) and variance -- but can also be explored through other methods such as flickering and skewness \citep{dakos2013}, via direct, regression-based estimates of the recovery rate as a proxy of stability \citep{boers2021b,smith2022}, via spectral properties \citep{bury2020,morr2024}, or using deep learning techniques \citep{bury2021,huang2024}. 

The recently introduced dynamic eigenvalue approach \citep{grziwotz2023} -- motivated by the estimation of the Jacobian around a locally linear time window -- directly quantifies stability; if any eigenvalue crosses 1, the system is unstable. The application of eigenvalues in the estimation of stability opens the door to new possibilities; in particular, it is possible to move beyond the analysis of a single time series and analyze spatio-temporal grids representing a spatially extended system state through time. In essence, it is possible to study the coherent evolution of an entire system forward in time, to understand whether the system as a whole is stable or not, and to capture the spatial patterns associated with that (in)stability.

In this work, we first build upon the recently proposed eigenvalue approach to estimate CSD \citep{grziwotz2023} on single time series with a flexible and generalizable methodology based on Dynamic Mode Decomposition (DMD) \citep{schmid2022}. We further show how our method can be adapted to natively handle periodicity via the calculation of Floquet Multipliers \citep{floquet1883}, hence removing a key source of potential bias due to uncertainty in pre-processing methodologies \citep{smith2023b}. We illustrate the utility of this new method using a glacier system with highly variable seasonal velocities that are difficult to decompose with typical pre-processing methods; in such cases, the application of common CSD indicators (AC1, variance) is not straightforward. Finally, we show how our proposed methodology can also be extended spatially, which opens the door to the study of a much wider range of systems and allows for the spatial patterns of destabilization to be analyzed through the tracking of spatial modes associated with unstable eigenvalues. We illustrate this approach for glacier surging as well as for the widely discussed potential destabilization of the Amazon rainforest \citep{boulton2022,lenton2024}. We emphasize that our new methodology is flexible to different data types and resolutions, and enables the study of a much larger set of climatic, geomorphic, cryospheric, and environmental processes than have so far been studied under the umbrella of CSD.  

\section*{Estimating the Stability of Seasonal Systems}

The core assumption underlying the application of typical CSD approaches is that the system is stationary, i.e. that there are no trends or expressions of seasonality that might bias the CSD indicators. In practice, this means that many types of data -- for example, vegetation indices -- need to be detrended and deseasoned before their stability can be assessed within a CSD framework \citep{smith2023b}. While linear detrending is often straightforward, it can become arbitrarily complicated in the case of nonlinear trends such as those imposed on the climate system by slow modes of natural variability; the process of removing seasonality is even more error-prone \citep{smith2023b}. A wide range of approaches have been used in the literature, including subtracting long-term climatological means \citep{forzieri2022}, fitting harmonic functions \citep{verbesselt2016}, seasonal trend decomposition via Loess (STL\citep{STL}), and simple rolling averages over full seasonal periods \citep{smith2023b}. While all of these approaches can remove seasonality and yield a nominally stationary residual time series, it is not always clear which parts of the signal should be considered noise, which parts seasonality, and which parts trend. In essence, the process of detrending and deseasoning data can lead to spurious signals which may be misinterpreted as stability changes or over-/under-estimation of resilience \citep{smith2023b}. Furthermore, when considering a set of time series (e.g., vegetation indices over a large region), each time series must be deseasoned and detrended individually, potentially introducing additional artifacts due to differences in the removed seasonality of each time series.

\begin{figure*}[!h]
\centering
\includegraphics[width=0.75\linewidth]{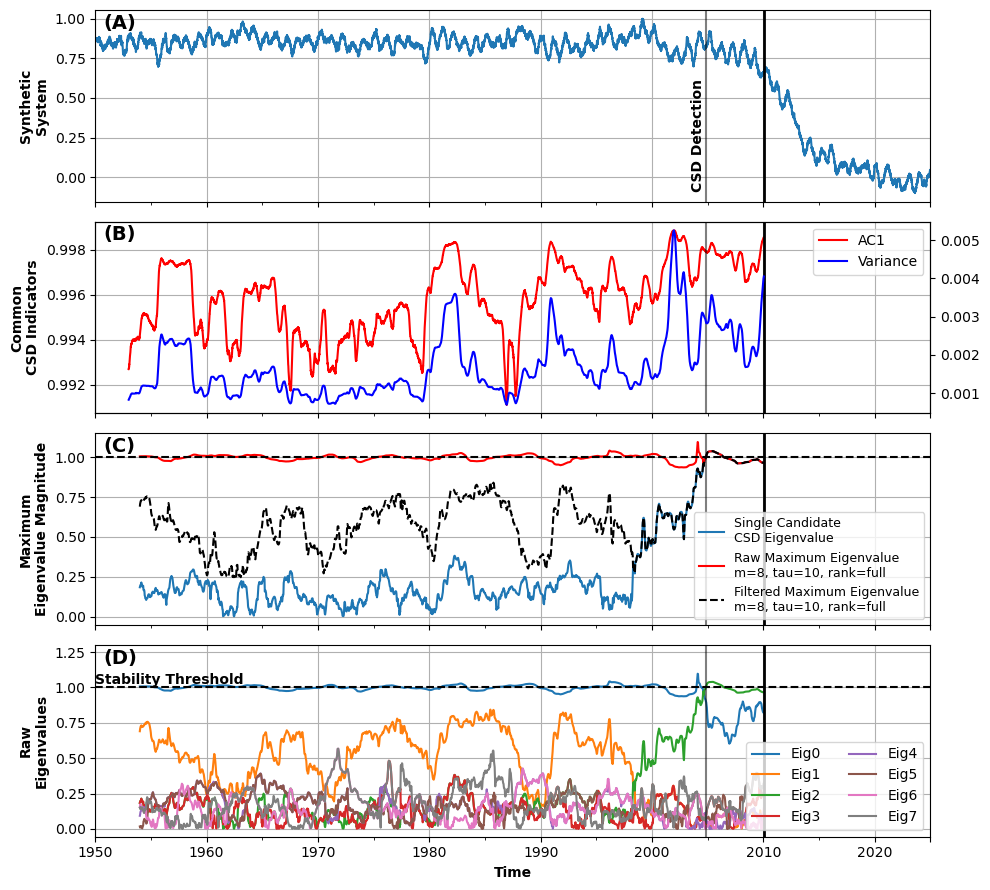}
\caption{\textbf{Floquet Multipliers}. (A) Simple model with seasonality moving towards a state transition. Vertical line shows instability detection from eigenvalue tracking. (B) Lag-one autocorrelation and variance computed directly on the time series without deseasoning. (C) Filtered eigenvalues, showing that a dominant eigenvalue approaches and crosses 1 before the critical transition. There is also a single eigenvalue that remains close to 1, which captures the stability in the periodicity of the time series (red line). (D) The set of all eigenvalues shows that most of them capture short-term noise in the system rather than coherent changes.}
\label{f1}
\end{figure*}

It is possible, however, to estimate the stability of a dynamical system without enforcing stationarity, for example through analysis of changes in periodicity or the dynamics superimposed on a regular periodic cycle \citep{floquet1883}. One approach is to compute the Monodromy matrix, which is solved over the known or computed periodicity of a given system. The eigenvalues of that matrix are termed Floquet Multipliers, and quantify stability around a system's periodic orbit \citep{floquet1883,coddington1955,guckenheimer2013}. In practice, this approach provides a framework for quantifying CSD-based resilience or stability in periodic data without first having to remove the periodicity; hence, ambiguity around deseasoning methodologies is removed. We test this assumption using a simple model with seasonality moving towards a critical transition (Figure \ref{f1}). 

The use of AC1 and variance on non-stationary time series is not supported by the underlying theory of CSD \citep{Djikstra2013,boers2022_erl}; as can be seen in Figure \ref{f1}b, both are strongly influenced by seasonal fluctuations in the time series, and cannot be used as a warning of a critical transition without first being deseasoned. Floquet Multipliers, however, can capture changes in the underlying driving process around a stable periodicity (Figure \ref{f1}c). In short, a stable periodic component (or stable trend) is captured by a single strong eigenvalue with a magnitude that remains close to 1, while the process driving the system towards a critical transition is captured by a secondary eigenvalue (or the maximum of all non-periodic eigenvalues). Further eigenvalues capture short-term noise and are relatively low-magnitude (Figure \ref{f1}d). The estimation of Floquet Multipliers (Methods) thus allows for the analysis of seasonal data within a CSD framework, without first resorting to deseasoning. 

Recent work has identified one further key limitation to using the standard CSD indicators AC1 and variance in many cases: multi-instrument data \citep{smith2022b}. It is often the case that instrumentation upgrades change the fidelity or dynamic range of measurements. These changes -- for example, modified signal-to-noise ratios -- leave traces in especially the variance of a time series, but also autocorrelation \citep{smith2022b}. This can make it difficult to differentiate changes in e.g. the variance driven by a loss of stability and those driven by measurement changes. In this sense, our approach has a key advantage compared to standard CSD indicators: it is less sensitive to changes in noise levels along a time series (Supplemental Figure S1). 

While DMD is not immune to noise, it is designed to extract coherent and continuous signals from noisy data; in essence, it attempts to reconstruct the most important parts of the underlying signal despite noise. It accomplishes this through rank truncation (Methods), which is a powerful tool for eliminating short-term and incoherent signals while preserving the main driving process; there also exist several extensions to the basic DMD algorithm designed to handle noise-induced biases \citep{dawson2016}. By reconstructing the same data used in Figure \ref{f1} with variable noise levels (Supplemental Figures S2-S4), we show that noise-level changes within reasonable ranges do not induce strong biases in eigenvalue-based CSD detection as they do in the case of AC1 and variance (Methods). 

\subsection*{Predicting Glacier Surging}

Recent advances in image cross correlation have yielded global coverage of glacier velocities at high temporal frequency \citep{gardner2018,gardner2025}. Glaciers are to some degree seasonal -- they build mass during the accumulation (e.g., winter) season and lose mass during the ablation (e.g., summer) season. The speed at which glaciers move is roughly proportional to their mass; heavier glaciers tend to move faster \citep{dehecq2019}, given similar slope and bedrock conditions. In many glaciers, downslope movement is not regular, but rather moves in distinct phases. At low mass loadings, glaciers can become `locked' and move very slowly or not at all. At some critical threshold, many glaciers can start to move very quickly in what is known as a glacier surge. While the exact conditions that trigger such surges are still debated \citep{ou2022,kaab2023,guillet2022,dehecq2019,benn2019,benn2023,thogersen2019}, many glacier regions worldwide are known to have periodically surging glaciers. 

\begin{figure*}[!h]
\centering
\includegraphics[width=\linewidth]{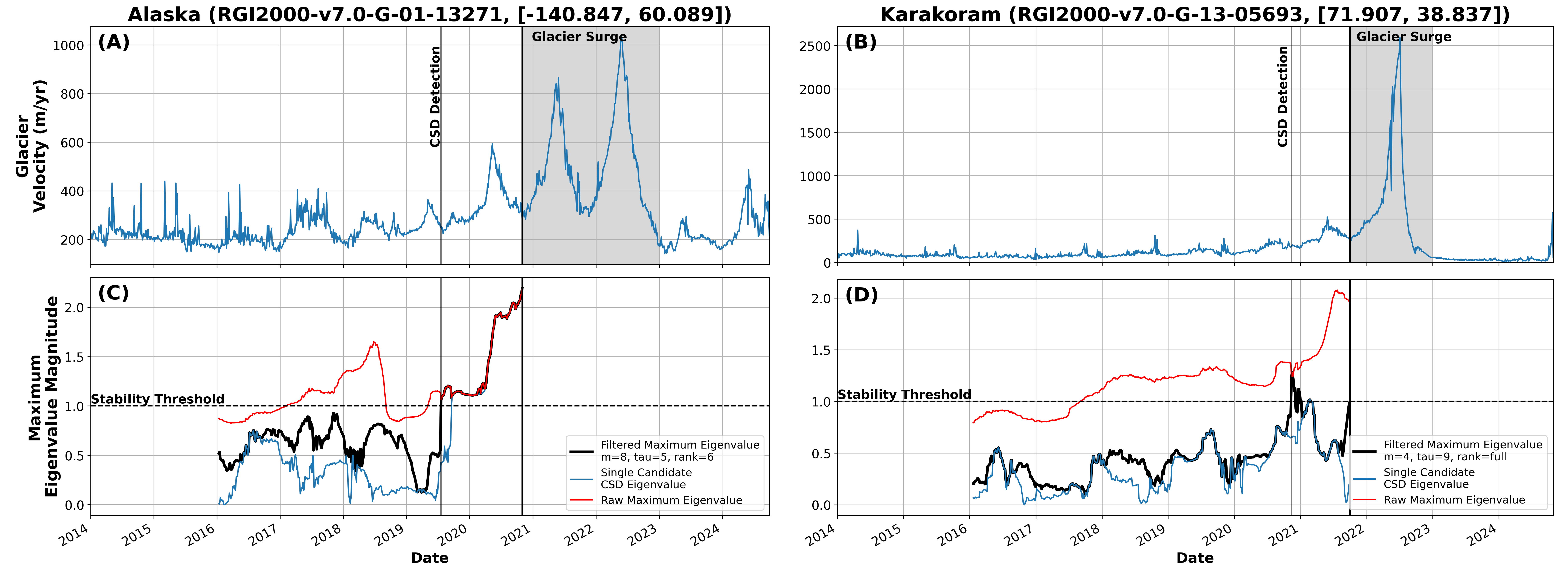}
\caption{\textbf{Glacier Instability}. Left panels: Glacier 1 in Alaska (RGI2000-v7.0-G-01-13271, [-140.847, 60.089]). Right panels: Glacier 2 in the Karakoram (RGI2000-v7.0-G-13-05693, [71.907, 38.837]). (A,B) velocity data resampled to 4-day consensus mean values (Methods). Shaded area shows surge period based on velocity magnitude. Eigenvalue-based CSD glacier surge detection marked with a vertical line. (C,D) Dominant eigenvalues using a period-lagged moving window (period = one year). Despite drastic differences in the type of glacier (coastal and mountain) and the relative magnitude of `fast' and `slow' periods, instabilities are both clearly marked by an eigenvalue (or the filtered maximum) crossing the stability threshold (magnitude \textgreater 1) before the surge.}
\label{f2}
\end{figure*}

Unlike vegetation data, which has been studied extensively using CSD \citep{smith2022,forzieri2022,boulton2022}, glacier velocity data is ill-suited to deseasoning techniques. The variability in seasonal velocity amplitudes can be several orders of magnitude; typical long-term mean or rolling-window smoothed estimates of seasonality, hence, leave large traces in the nominally deseasoned and stationary time series as required for using AC1 and variance to quantify stability and its changes within the CSD framework. To overcome this challenge, we estimate Floquet Multipliers across a set of moving windows (Methods) instead of typical AC1 and variance estimates. We use velocity time series at single points over the period 2014-2025 for two test glaciers, one in Alaska (RGI2000-v7.0-G-01-13271, [-140.847, 60.089]) and one in the Karakoram (RGI2000-v7.0-G-13-05693, [71.907, 38.837]) \citep{rgi} (Figure \ref{f2}).

Seasonality in both glaciers varies significantly through time (Figure \ref{f2}); the high and increasing magnitude of the dominant eigenvalue is a clear sign of significant and growing instability in that periodicity for both glaciers. Beyond changes in the seasonal cycle, however, both glacier examples show further destabilization in the non-seasonal parts of the signal, with a lead time of at least one year before surging (Figure \ref{f2}c,d). 

This hints at two possible ways to use Floquet Multipliers in the context of glacier changes: (1) look for strong and consistent increases in the largest/most strongly periodic eigenvalue to quantify destabilization in the seasonal cycle, and (2) examine whether additional eigenvalues are moving towards 1 to indicate non-periodic instabilities related to potential state changes (i.e., glacier surges). Both predictions could be of great practical use, even if they can only forecast the onset of glacier surging one or two years in advance. As glacier surges are key to understanding glacier contributions to sea-level rise \citep{immerzeel2010} and predicting glacier lake outburst flood risk \citep{veh2023}, even short warning windows are invaluable.  

\section*{Beyond Time Series: Tracking the Stability of the Whole System}

Single time series are often used to estimate the stability of a system and changes thereof, serving as low-dimensional proxies for the behavior of large, spatially extended systems \citep{boers2021b,grziwotz2023}. This simplification is motivated by the ability of a single state variable to reconstruct a wider state-space, often using high-dimensional embedding (e.g., Takens's Theorem \citep{takens1981}). Alternatively, for systems where spatial patterns are of interest, many time series can be analyzed in parallel (e.g., global vegetation resilience maps \citep{smith2022,forzieri2022}). One of the main motivations for this approach is that deseasoning and detrending are only feasible on individual time series and not regions as a whole; the seasonality of spatially adjacent time series is not guaranteed to be identical. 

If we instead consider a spatial area -- for example, a single glacier -- as a coherent system, we can use spatio-temporal data to understand the development of the system as a whole. Applying the same mathematical framework of DMD eigenvalue tracking and Floquet Multipliers to spatio-temporal snapshots of a given system yields three main advantages: (1) we have a denser and more accurate view of the whole system state; (2) we can visualize which spatial modes contribute to any CSD eigenvalues, and hence understand the spatial patterns of destabilization; and (3) spatially uncorrelated noise will be broadly removed, as it will be expressed as rapidly decaying modes with low eigenvalue magnitudes. We test the utility of this method on a system with a known transient state (a glacier which surges, cf. Figure \ref{f2}) and one that has been proposed to be moving towards a critical threshold, i.e. the Amazon rainforest \citep{boulton2022,lenton2024}. 

\subsection*{Glacier Surging}

\begin{figure*}[!h]
\centering
\includegraphics[width=\linewidth]{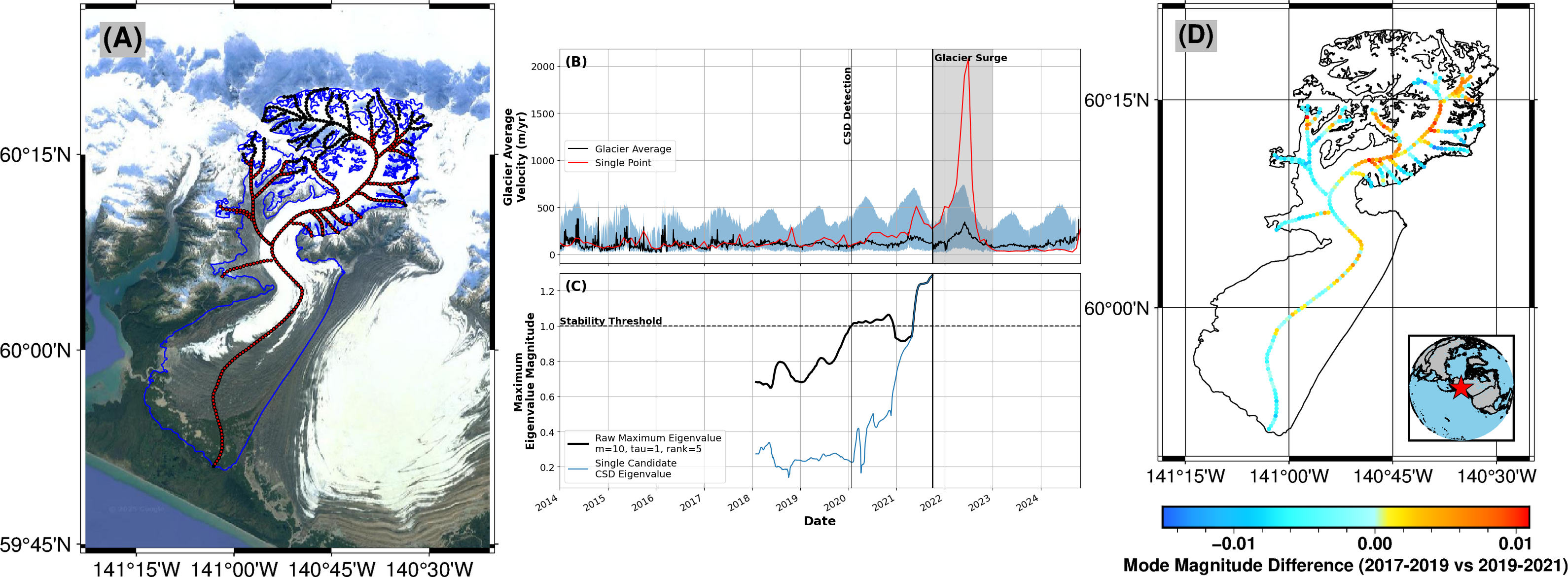}
\caption{\textbf{Whole-Glacier Stability}. (A) Map of the target glacier in Alaska (RGI2000-v7.0-G-01-13271, [-140.847,60.089]) with outline (blue) and sample points ($n=337$) along the glacier centerline (red, low-elevation points, black, all centerline points). Background from Google Earth. (B) 4-day ($t=1004$ time steps) glacier-average velocity (black) plus 25th-75th percentile range (shaded blue), alongside single strongly surging point (red). Shaded area shows surge period based on velocity magnitude. Eigenvalue-based CSD glacier surge detection marked with vertical line.  (C) Maximum eigenvalue magnitude using the set of velocities over all low-elevation centerline points, showing a trend towards instability far before the actual surge. (D) Change in the spatial mode (Methods) between stable and directly pre-surge periods. The change indicates that the increase in eigenvalue magnitude is primarily driven by locally activated sections of the glacier. Locator map as inset in bottom right.}
\label{f3}
\end{figure*}

One-dimensional data (e.g., velocity time series, Figure \ref{f2}) is often a drastic simplification of a system under study. From a mathematical perspective, the linear operator that maps a system state forward in time (Methods) can also be estimated from the entire spatio-temporal data cube that describes that system -- in essence, the snapshot estimates of velocity over the entire glacier surface can be used instead of velocity measurements taken at a single selected point on the glacier (Figure \ref{f3}). While we use velocities measured along the glacier centerline here for simplicity, gridded velocity estimates over the entire glacier could also be used in the same way; this adds computational complexity but could further refine the spatial patterns of glacier surge initiation. We use three-year moving windows for computational stability and period-lag by one year for the inferred periodicity of glacier movement in order to capture Floquet Multipliers (Figure \ref{f3}).

The snapshot spatial field of velocity estimates is already an improved measurement of the system state, given that it contains multiple velocity measurements alongside location information rather than just velocity for a single point location. If that pairing (velocity, location) was enough to fully describe how a glacier evolves, it would imply that the stability of the glacier as a whole could be well-predicted from only its current and period-lagged velocity fields. In our tests, this did, in fact, work to predict glacier surging (Supplemental Figure S5), but adding additional embedding dimensions (i.e. time lags, Methods) gave earlier and stronger warning signals of impending surging; in short, the stability of a glacier is better described by location, velocity, and acceleration over multiple time periods (Figure \ref{f3}). In principle, more embedding dimensions can be used; this increases the computational cost and risk of overfitting, but can capture longer wavelength changes in the glacier. In our tests, low embedding dimensions ($m \leq 3$) were sufficient to pick out the signs of an impending surge using the well-monitored period of 2014-2025 (Supplemental Figure S5), but a higher ($m=10$) embedding dimension picked out the oncoming surge earlier. We also tested several different time samplings of the glacier velocity field -- between four and 30 days -- and found qualitatively similar results (Supplemental Figure S6).

When we compare the timing of surge initiation in Figure \ref{f3} to that found in Figure \ref{f2}a, there is a clear difference -- the 1D case seems to spot the onset of a surge earlier. It requires, however, that we choose a point on the glacier that is of interest -- a choice that is not always easy. By looking at the whole glacial system, we can instead examine the overall stability of the glacier, without focusing on a specific point, and without possible problems associated with the representativeness of that point for the entire glacier. This also allows us to, for example, look at the spatial modes associated with each eigenvalue through time to see which parts of the glacier are contributing to the overall (in)stability of the system (Figure \ref{f3}d). From this, we can see that the surge activity is concentrated in distinct areas; it is not expressed as an acceleration evenly throughout the glacier, but rather in spatial subsets that change through time. This illustrates that we capture the growing instability in the glacier despite the spatially disconnected nature of surge initiation.

\subsection*{The Amazon Rainforest}

\begin{figure*}[!h]
\centering
\includegraphics[width=\linewidth]{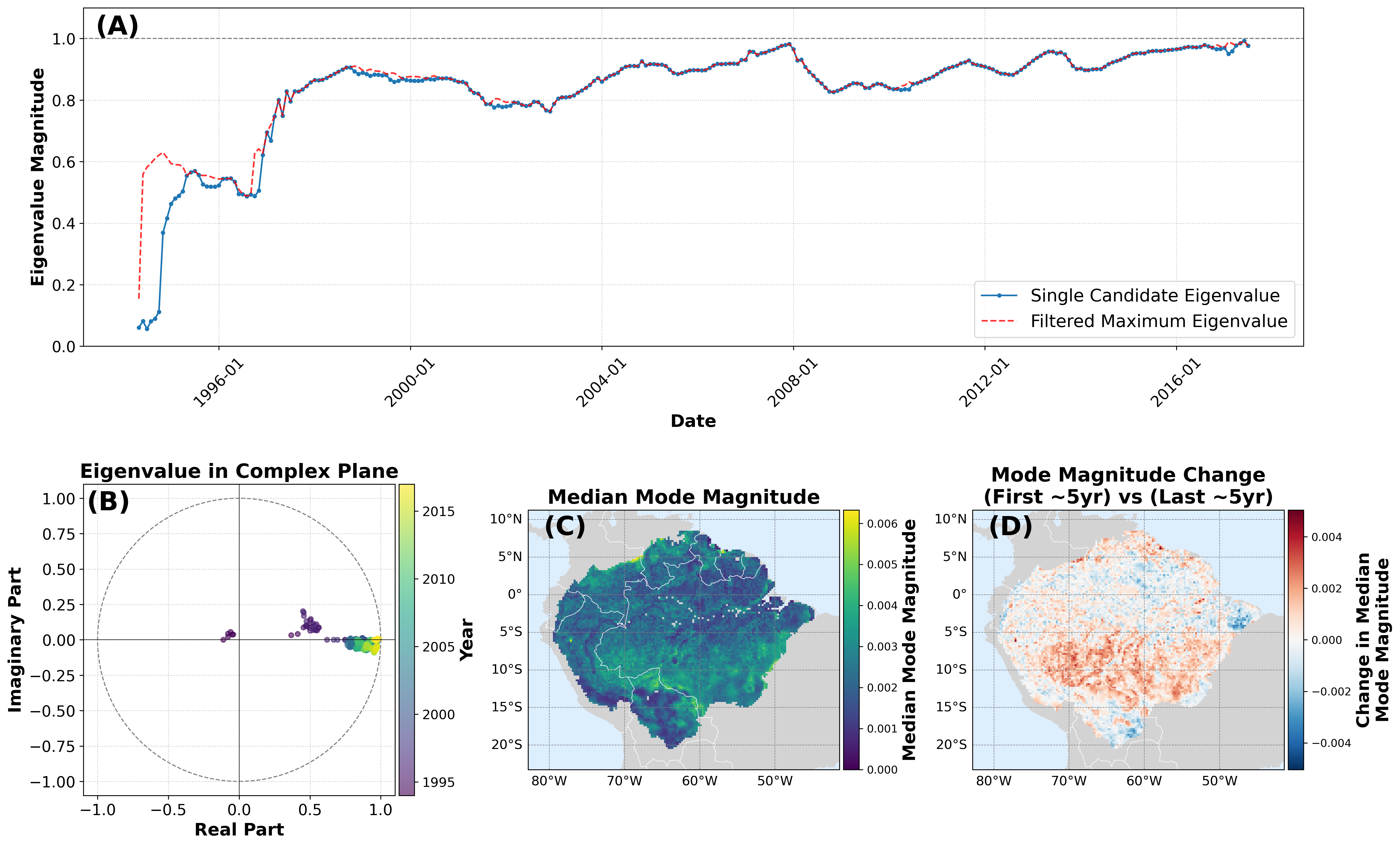}
\caption{\textbf{Amazon Vegetation Stability}. (A) Single eigenvalue of interest and maximum of all non-periodic eigenvalues. (B) Eigenvalue track in the complex plane, showing a movement towards the unit circle over time, indicating loss of stability. (C) Spatial map of the median spatial mode associated with the eigenvalue track of interest, showing that the mode is most strongly associated with the southern parts of the Amazon basin, which have been significantly affected by deforestation in recent decades \citep{hansen2013}. (D) Change in the spatial mode map between the first and last five years of data. The change indicates that the increase in eigenvalue magnitude is strongest in the southern Amazon, as well as some areas on the northern edge of the Amazon.}
\label{f4}
\end{figure*}

The ability to track the spatial patterns of destabilization in a coherent system is a key advantage of using spatio-temporal grids instead of 1D time series. To further illustrate this point, we use the widely-studied example of vegetation dynamics in the Amazon rainforest (Figure \ref{f4}). A growing body of work has proposed that the Amazon rainforest could potentially tip into a savanna-like state due to a breakdown of moisture recycling, which currently still sustains dense vegetation \citep{blaschke2024}. Previous work by several authors has demonstrated that common CSD metrics -- for example, AC1 and variance -- indicate that the Amazon has been losing resilience in recent decades \citep{boulton2022, smith2022, blaschke2024}; however, the timing and spatial scale of that critical transition remain unclear. 

Using monthly-averaged vegetation optical depth (VOD) estimates \citep{VOD}, we can assess the stability of the Amazon as an entire system, rather than relying on the spatial pattern of individually analyzed time series (Figure \ref{f4}). The spatial mode associated with the eigenvalue of interest -- that which is approaching 1 from below -- illustrates which part of the system contributes most strongly to that destabilizing process (Figure \ref{f4}c); looking at changes in that spatial mode through time also gives insight into which parts of the system are contributing most strongly to the overall destabilization (Figure \ref{f4}d). 

The tracked eigenvalue in Figure \ref{f4}a has the highest mode magnitude in the southern Amazon (Figure \ref{f4}c), which has seen extensive deforestation in recent decades \citep{hansen2013}. It is thus not entirely surprising that this area behaves differently from the rest of the Amazon, and that it forms a distinct spatial mode in Amazon vegetation changes (Figure \ref{f4}d). It should be noted, however, that even a stable seasonal mode (Supplemental Figure S7) also captures a divide in vegetation dynamics between the internal and external portions of the Amazon. The stability of that strongly periodic mode implies that the \textit{seasonality} of the Amazon is not changing drastically (as it is, for example, for the glaciers shown in Figure \ref{f2}), but rather that some other non-periodic change is leading to reduced stability in the Amazon rainforest. 

\section*{Discussion}

There is a growing interest across scientific disciplines in incorporating dynamical systems frameworks and especially the CSD concept into research concerned with stability changes in natural systems \citep{dakos2019,van2024,blom2024,lenton2024,tang2022}. At the same time, several recent publications have underlined critical biases in the most commonly used CSD-based stability indicators -- variance and AC1 \citep{boers2021b,smith2022b, smith2023b, grziwotz2023}; extending the study of system stability to use the well-motivated eigenvalue magnitude approach \citep{grziwotz2023} is a welcome addition. By extending this new eigenvalue-based method of resilience estimation to new domains -- both seasonal and spatio-temporal data -- the number and diversity of systems that can be studied within a CSD framework increases dramatically.

The main limitation to the use of eigenvalue tracking and Floquet Multipliers in the study of system stability is the choice and availability of data. The example of glacier surging shown here is one with a clear transient state (surging), which is expressed directly and completely in the state variable measured (velocity), which makes for a relatively easy system to study with our method. Landslides behave in some similar ways to glaciers; mass loading leads to downhill pressure, stabilized by the cohesion of the soil and friction with the underlying bedrock. The triggering mechanism for a landslide, however, is not as simple as that for a glacier surge; external factors such as rainfall, water table levels, earthquakes, and land-cover changes also play important roles. It is thus not clear if the data available -- GPS monitoring, radar interferometry, pixel tracking velocity fields -- is really capturing the system state well enough to predict a critical shift in stability. This same data limitation applies to other Earth system components that could be considered as unstable or multi-state dynamical systems, for example earthquakes (plate motion adjustment) and volcanoes (release of magma pressure). The directness of the measurement data -- and complications due to constant adjustments to small earthquakes -- make the direct application of our analysis framework to those systems difficult. 

Remote sensing, however, provides a powerful tool that is perfectly suited to the use of DMD for stability monitoring; indeed, DMD was initially developed for spatio-temporal grids \citep{schmid2010,kutz2016,schmid2022}, and has been used to discover dynamical patterns in climate systems \citep{lapo2025}. We demonstrate how DMD can be used with remote sensing data through our analysis of glacier (Figures \ref{f2}, \ref{f3}) and Amazon rainforest stability (Figure \ref{f4}). This same framework could be used, for example, to study ocean circulation systems \citep{boers2021b}, river avulsion \citep{gearon2024}, changes in ice export in the north Atlantic \citep{smedsrud2017}, or spatial vegetation pattern formation \citep{donovan2022,wang2022}. We emphasize that the mathematical framework we present here is domain-agnostic and can be applied equally well to 1D, 2D, or higher-dimensional data. 

Our work builds upon recent theoretical advances in resilience estimation with eigenvalue approaches \citep{grziwotz2023} to present a novel method that inherently handles seasonality and can also be natively extended to high-dimensional data and spatio-temporal fields. We present examples from synthetic, cryospheric, and ecologic systems that are thought to have multiple stable or transient states, demonstrating the flexibility and usefulness of our method. It is well-suited to diverse application areas, and to take advantage of the growing amount of multi-scale data available across diverse spatial and temporal resolutions. 

\clearpage
\newpage

\section*{Methods}

\subsection*{Eigenvalue Estimates of Linear Stability}

Recent work \citep{grziwotz2023,kulkarni2024} has proposed the use of dynamic eigenvalues as an early warning of critical transitions. This use is motivated by locally linearizing a system around a stable point and calculating the Jacobian. For a stable system, all discrete-time eigenvalues of the Jacobian should be real and less than 1; when any eigenvalue approaches or crosses 1, the system becomes unstable. There are many approaches to estimating the Jacobian of a local linearization. Grziwotz et al. \citep{grziwotz2023} use an S-Map \citep{cenci2019,deyle2016} approach based on a local nearest-neighbors search at every point in a time series. By moving windows across a time series and averaging the dominant (largest) eigenvalue over each window, the stability of the system can be estimated through time. 

Dynamic Mode Decomposition (DMD) provides an alternative approach to estimating the Jacobian \citep{schmid2010,kutz2016,schmid2022}. DMD takes snapshot matrices representing subsequent system states and uses singular value decomposition (SVD) to compute a single matrix $\mathbf{A}$ that provides the mapping $\mathbf{x}_{k+1} \approx \mathbf{A} \mathbf{x}_k$. The eigenvalues of that linear operator $\mathbf{A}$ are a data-driven approximation of the underlying driving process of the system, and provide an alternative means of estimating the stability of the system; in short, DMD finds a set of eigenvalues for a global fit to the system state in one (temporal) window, while S-Map as applied by \citep{grziwotz2023} finds point-wise local eigenvalues and then averages them over a given (temporal) window. For a simple dynamical system with a bifurcation, the two methods yield similar results (Supplemental Figure S8). This can also be demonstrated with real-world data that does not have such simple dynamics (Supplemental Figure S9), but was previously shown to be approaching a tipping point \citep{boers2021}. 

\subsubsection*{Using DMD to Compute Floquet Multipliers}

Many natural systems have an inherent periodicity, most often due to daily or annual cycles related to sunlight or temperature. The analysis of such systems within the CSD framework has so far focused on using deseasoned and detrended data, introducing potential biases in resilience estimates from pre-processing methodologies \citep[e.g.][]{verbesselt2016, smith2022b, smith2023b}. It is, however, possible to estimate changes in linear stability over a given periodicity and hence, estimate resilience within a CSD framework without having to first remove seasonality. 

Consider a linear system with a periodic coefficient matrix $\mathbf{A}(t)$:

\begin{equation}
\frac{d\mathbf{z}}{dt} = \mathbf{A}(t) \mathbf{z}, \quad \text{where } \mathbf{A}(t+T) = \mathbf{A}(t)
\end{equation}

\noindent where $T$ is the known period of the system and $\mathbf{z}$ are perturbations around the periodicity. The fundamental matrix solution is $\mathbf{\Phi}(t, t_0)$, with the state of the system evolving as $\mathbf{z}(t) = \mathbf{\Phi}(t, t_0) \mathbf{z}(t_0)$. The Monodromy matrix $\mathbf{M}$ captures how the system evolves over one full period $T$ as $\mathbf{M} = \mathbf{\Phi}(t_0+T, t_0)$. The eigenvalues $\{\rho_i\}$ of $\mathbf{M}$ are the Floquet Multipliers \citep{floquet1883,coddington1955,guckenheimer2013}, which describe the linear stability of the system over the period $T$. For stable systems, $|\rho_i|$ will be less than 1 for each eigenvalue $\{\rho_i\}$; any eigenvalue \textgreater 1 implies destabilization in the system. 

To compute Floquet Multipliers with DMD, we rely on the link between DMD and Koopman theory \citep{koopman1931,rowley2009}. Koopman theory allows us to reformulate a non-linear, finite-dimensional system as a linear, infinite-dimensional system, where the infinite-dimensional space is spanned by so-called observables $\phi$. The dynamics of each observable of the system can thus be written as:

\begin{equation}
\phi(t_{k+1})= \mathcal{K}^T (\phi(t_k))
\end{equation}

\noindent where $\mathcal{K}^T$ is the linear Koopman operator over the period $T$ \citep{koopman1931,rowley2009}. For any state in the system near the periodic orbit, $\mathcal{K}^T$ is an approximation of the Monodromy matrix $\mathbf{M}$. If the initial system was linear to begin with, these operators are identical. In practice, for snapshots separated by $T$, DMD can be used to find a finite-dimensional approximation of $\mathcal{K}^T$\citep{rowley2009}. This is achieved by computing a linear operator $\mathbf{A}_{T}$ in the original basis so that $\mathbf{y}_k \approx \mathbf{A}_{T} \mathbf{x}_k$, where $\mathbf{y}_k$ is the system state exactly one period $T$ after $\mathbf{x}_k$. Thus, the $\mathbf{A}_{T}$ computed with DMD is a data-driven estimate of the linear evolution of the system for one full period $T$. For periodic dynamics, this map is a data-driven approximation of the Monodromy matrix $\mathbf{M}$; hence the eigenvalues of $\mathbf{A}_{T}$ are approximations of the Floquet Multipliers $\{\rho_i\}$. Thus, we can estimate the stability of the underlying system without having to explicitly remove seasonality, and track eigenvalue changes \citep{grziwotz2023} to understand the stability of inherently periodic systems (Figure \ref{f1}).

The tracking of stability with Floquet Multipliers is complicated by the fact that an eigenvalue indicative of CSD will approach 1 from below, while a stable periodic oscillation (or stable linear trend or mean state) will have an eigenvalue of $\sim$1 \citep{Strogatz2015}. In systems with noise, it can become very difficult to separate eigenvalues representing drift towards a critical transition from those that are associated with stable periodicity. A simple maximum eigenvalue (as used by \citep{grziwotz2023} with S-Map) will capture small oscillations of that stable periodicity -- not the destabilization of the underlying linear operator. The stable periodic eigenvalue will have, in a perfect case, a value of [1 + 0j]; with noise, this can oscillate around 1 and also pick up small imaginary components. However, in general this periodic mode can be filtered out by removing eigenvalues that are stable and oscillate around 1 throughout the entire period of study, leaving a filtered maximum eigenvalue representing stability change; alternatively, a single candidate eigenvalue can be identified that approaches 1 from below. It is also important to note that mean-centering the windowed data can minimize the magnitude of the stable periodic eigenvalue (e.g., if the amplitude of the periodicity is small), and hence enhance the relative strength of potential CSD modes. We do not mean-center the data in Figures  \ref{f1} and \ref{f2} to illustrate that our method can capture both periodic and non-periodic signals; the impacts of mean-centering can be seen in Supplemental Figure S10. 

\subsubsection*{Implementation Details: Period Lagged DMD}

A core requirement for eigenvalue-tracking-based CSD analysis is that the local linear maps (estimated via S-Map, DMD, or other methods) are computed over a sufficiently dense picture of the underlying system. In practical terms, this means that time series data needs both temporal lags $\tau$ and embedding dimensions $m$ to fully capture the attractor. Takens's embedding theorem \citep{takens1981} provides a rule of thumb approximation for the choice of $\tau$ and $m$. In the examples shown here, $\tau$ is first estimated via average mutual information, choosing the first local minima over successive lags from 1 to $T / 2$. Building from there, we use False Nearest Neighbors (FNN) \citep{kennel1992} to compute the ideal $m$ embedding dimension. In noisy systems, these are not guaranteed to be the ideal $\tau$ and $m$; Takens's theorem is meant to reconstruct the attractor over infinite and noise-free data, and can fail to find the optimal $m,\tau$ parameters in real noisy systems. For our toy systems, we assess a range of $\tau$ and $m$ parameters close to the initial guesses from mutual information and FNN, choosing the pair with the lowest $m$ that is able to separate the CSD signal mostly into its own eigenvalue. For the simple case without seasonality, we use a straightforward $m=1$ setup. The resulting eigenvalue changes found by S-Map and DMD are very similar (Supplemental Figure S8).

The choice of embedding dimension $m$ and lag $\tau$ for seasonal data via Floquet analysis (Figure \ref{f1}) is more difficult due to the confounding influence of the seasonal eigenvalue (close to 1) on the detection of the CSD eigenvalue (below but approaching 1). Further, average mutual information and FNN are designed to find linearized dynamics, not period-lagged dynamics; for our periodic toy model, they generally reconstruct the strong seasonal attractor and, to some degree, ignore the underlying CSD signal of interest. To overcome these limitations, we test a range of $m/\tau$ values, choosing a pair that cleanly separates out the CSD eigenvalue of interest; the choice of $m/\tau$ will depend heavily on the system under study, the driving noise, and the memory time scale of the system.

Once a set of $m,\tau$ parameters has been chosen, as well as the DMD rank truncation $r$ and period $T$, the data is processed as a set of overlapping windows. For each window, a set of delay-embedded matrices is formed such that a single time point is an $m$-dimensional vector: 

\[
\mathbf{x}_t = \begin{pmatrix}
x_{t} \\
x_{t+\tau} \\
x_{t+2\tau} \\
\vdots \\
x_{t+(m-1)\tau}
\end{pmatrix}
\]

where $x_{t}$ is the value of the time series at time $t$. A single window of data $\mathbf{X}$ of length $n$ is an $m \times n$ matrix:

\[
\mathbf{X} = \begin{pmatrix}
x_{t} & x_{t+1} & x_{t+2} & \dots & x_{t+n-1} \\
x_{t+\tau} & x_{t+1+\tau} & x_{t+2+\tau} & \dots & x_{t+n-1+\tau} \\
x_{t+2\tau} & x_{t+1+2\tau} & x_{t+2+2\tau} & \dots & x_{t+n-1+2\tau} \\
\vdots & \vdots & \vdots & \ddots & \vdots \\
x_{t+(m-1)\tau} & x_{t+1+(m-1)\tau} & x_{t+2+(m-1)\tau} & \dots & x_{t+n-1+(m-1)\tau}
\end{pmatrix}
\]

This is compared to a period-lagged matrix $\mathbf{Y}$:

\[
\mathbf{Y} = \begin{pmatrix}
y_{t+T} & y_{t+T+1} & y_{t+T+2} & \dots & y_{t+T+n-1} \\
y_{t+T+\tau} & y_{t+T+1+\tau} & y_{t+T+2+\tau} & \dots & y_{t+T+n-1+\tau} \\
y_{t+T+2\tau} & y_{t+T+1+2\tau} & y_{t+T+2+2\tau} & \dots & y_{t+T+n-1+2\tau} \\
\vdots & \vdots & \vdots & \ddots & \vdots \\
y_{t+T+(m-1)\tau} & y_{t+T+1+(m-1)\tau} & y_{t+T+2+(m-1)\tau} & \dots & y_{t+T+n-1+(m-1)\tau}
\end{pmatrix}
\]

The linear operator $\mathbf{A}$ is found via $\mathbf{A} = \mathbf{Y} \mathbf{X}^\dagger$, where $\mathbf{X}^\dagger$ is the Moore-Penrose pseudo-inverse of $\mathbf{X}$. DMD computes this via SVD with optional rank truncation $r$, yielding either the full operator $\mathbf{A}$ or a reduced operator $\tilde{\mathbf{A}}$. In this specific period-lagged setup, the eigenvalues of $\mathbf{A}$ or $\tilde{\mathbf{A}}$ are interpreted as Floquet Multipliers. 

\subsubsection*{Application to Spatio-Temporal Data}

The estimation of eigenvalues via DMD is not limited to the 1D case; a very wide body of work uses DMD to analyze spatio-temporal data \citep{schmid2010,kutz2016,schmid2022,lapo2025}. In our case, the underlying motivation does not change between the 1D and 2D cases -- we are still estimating a linear operator $\mathbf{A}$ that moves the system state forward in time. The change is that we are no longer looking at a single state variable that captures the dynamics of the system, but rather focus on a \textit{collection} of state variables that change through time. While we use spatial fields of single parameters (e.g., velocity), there is no strong reason why multi-dimensional data (e.g., multiple satellite bands) could not also be used in the same way; the underlying theory of using period-lagged data for the estimation of Floquet Multipliers would not change, but the construction of the state vectors $\mathbf{X}$ and $\mathbf{Y}$ would have to be modified to accommodate multi-dimensional data. It is important to note that this approach would also work with non-periodic data (e.g., Supplemental Figures S8, S9) which has multiple measurement dimensions (e.g., GPS displacements in x/y/z directions).

One of the largest difficulties in using DMD -- particularly period-lagged DMD for Floquet multiplier estimation -- is the selection of $\tau$-lag and $m$-dimension embedding to reconstruct the underlying attractor \citep{takens1981}, and hence to capture the eigenvalue that is related to the stability of the system. As discussed above, while there exist rules of thumb for choosing the best $\tau$ and $m$ parameters for a given system (e.g., Takens's Theorem \citep{takens1981,packard1980}), in practice -- with noisy data -- these methods often fail. For spatio-temporal systems, there are also data processing constraints, necessitating a smaller SVD rank truncation. The maximum number of eigenvalues computed by DMD is limited by either the spatial field ($n$ features) or time ($t$ snapshots) by the minimum of $n-1$ and $t$ (the 1D case uses $m$ embedding dimensions instead of $n$ spatial locations). Depending on the system, this can generate a massive number of potential eigenvalues, many of which are short-time and decaying features or spatially uncorrelated noise. For example, strong noise in a single time series in a spatial field will not show up coherently throughout all time series, and would be unlikely to show up as a dominant eigenvalue. To rather focus on the main features of interest, we truncate DMD over each moving window and only preserve the dominant eigenvalue(s), letting quickly decaying modes fall away. For each spatio-temporal system studied here (Figures \ref{f3}, \ref{f4}), we test a range of $m$, $\tau$, and $r$ rank parameters, with a focus on cleanly separating periodic dynamics from non-periodic dynamics.

In the context of spatio-temporal fields, the structure of the dynamic modes (eigenvectors) associated with any given eigenvalue can be used to further investigate changes in the system. The spatial distribution of the dynamic mode magnitude -- especially that associated with a CSD-like eigenvalue -- can provide insight into which specific regions contribute most strongly to destabilization, and how their spatial pattern has evolved through time (Figures \ref{f3}, \ref{f4}). The dynamic modes are a natural output of DMD alongside the eigenvalues used to track stability here; when time-delay embedding is employed, further processing is needed to extract a single spatial mode from the embedded vector. To investigate only the influence of the current system state on the dynamics, we project the embedded modes back onto their first spatial component vector -- only the first $x$-vector of the $m$-embedded spatial $x$-vectors is analyzed. In short, delay embedding can be used to better reconstruct the state space and capture changes in the system; the spatial mode patterns are then removed from that state-space embedding to provide a single spatial mode per time window and eigenvalue. 

\subsection*{Toy Model Generation}

To compare standard (e.g., AC1) and novel (S-Map and DMD) methods for estimating stability changes in terms of CSD, we use simple toy models with a critical bifurcation (Figure \ref{f1}, Supplemental Figures S2-4, S8, S10):

\begin{equation}
dX_t = (-X_t^3 + X_t -p)dt + \sigma dW
\end{equation}

\noindent where $X_t$ is the system state at time $t$, $p$ is the control parameter that is varied to produce a state transition, and $W$ is a white-noise Wiener process with normally distributed increments driving the system. We produce two versions of this model: (1) with seasonality (Figure \ref{f1}, Supplemental Figures S2-4, S10) and (2) without (Supplemental Figure S8). We provide Python code to reproduce our synthetic time series here: \textit{link}. 

To test the impact of variable noise levels on DMD, as would for example arise by merging signals from different sensors \citep{smith2022b,VOD}, we perform two further experiments. We first generate a stationary process with additive seasonality, transform that time series into slices of variable length with different additional additive noise, and compute a daily average (Supplemental Figure S1). To generate a stationary residual for AC1 and variance calculation, we use STL \citep{STL}; we do not pre-process the daily signal for eigenvalue tracking. As a second experiment, we add time-varying and additive noise to the same model used in Figure \ref{f1}. We (1) increase, (2) decrease, and (3) shuffle the noise levels through time (Supplemental Figures S2-S4), in each case capturing changes in AC1 and variance on STL-processed residuals as well as eigenvalue changes on the data without pre-processing. We are able to successfully capture the signal of interest (eigenvalue approaching 1 from below) for the decreasing and shuffled noise cases. In the case of very high noise (noise magnitude \textgreater 2 times the signal magnitude) during the transition, the relatively weak CSD mode approaching 1 from below is hard to differentiate from noise, and can thus be missed. We emphasize, however, that this requires very high noise levels; in more typical real-world cases with rather smaller noise magnitudes and changes therein, our approach should remain robust. It is also important to point out that our method does not produce a spurious early-warning signal (as can happen with AC1 and variance), but rather does not find a CSD-like eigenvalue for very high noise levels.

\subsection*{Glacier Data}

Recent work has vastly expanded the amount and density of glacier velocity data available \citep{gardner2018,gardner2025}. We use ITS\_LIVE (version 2) velocity data sampled point-wise over the Landsat 8 era onward for two test glaciers: one in Alaska (RGI2000-v7.0-G-01-13271, [-140.847, 60.089]) and one in the Karakoram (RGI2000-v7.0-G-13-05693, [71.907, 38.837]) (Figure \ref{f2}) \citep{rgi}. These glaciers were chosen because they show two different modes of destabilization -- a multi-annual period of anomalously high velocity and a single surge period of extremely high velocity.

For each chosen glacier, we first select one single point in the center of the glacier tongue, and extract all available velocity measurements. Each measurement represents the average velocity over a defined period; we first assign each velocity measurement to the midpoint of those periods. To minimize the confounding influence on CSD estimation from different numbers of measurements through time \citep{smith2022b}, we use a consensus-building approach. In short, we first create a set of equally-spaced annual points (4-day) and assign all velocity measurements to one of those equally-spaced bins. We then take $n=100$ random iterations, choosing an equal number of points per bin based on the minimum number of contributing points across all bins. For example, if all equally-spaced bins have at least 5 points, we generate $n$ versions of our equally-spaced time series using a random sample of 5 points per time bin. We then average over the $n$ time series to create a single consensus time series of glacier velocity. In many cases, this is not significantly different from a rolling average; however, it explicitly accounts for the varying number of points per bin, so minimizes biases to the variance of the resulting time series.

In a second step, we generate a set of equally-spaced points along the glacier centerline at 500 m intervals. We further remove points in the upper portion of the glacier based on an underlying elevation model (Copernicus, 30 m) \citep{copdem}, with the assumption that the majority of the dynamics of interest are captured best in the lower reaches of each glacier. Since each point of the glacier is not guaranteed to have the same number of samples through time, we regularize the set of sampled points to an evenly-spaced time grid (4 days), leaving no-data values for missing points. This array of $n$ sample points through $t$ time steps is used directly with DMD without further pre-processing for the analysis shown in Figure \ref{f3}; the spatial pattern of velocities and missing data can be seen in Supplemental Figure S11. We note that we choose a small window (4 days) with the goal of \textit{maintaining} some of the noise rather than over-smoothing; noise local to a single time series is de-emphasized during the DMD processing, and thus we err on the side of keeping too much noise rather than removing too much, as can happen during, for example, typical deseasoning procedures \citep{smith2023b}. We confirm our results with a range of time samplings from 4 to 30 days (Supplemental Figure S6); in each case the same critical transition can be seen before surging. 

\subsection*{Vegetation Data}

We use monthly-averaged Ku-Band VOD data \citep{VOD} covering the broader Amazon basin over the period 1987-2017 (0.25 dd spatial resolution). VOD data has been shown to be superior to optically-based vegetation indices \citep{smith2022} for CSD analysis. We choose monthly-averaged rather than daily-scale data for ease of processing; our rolling-window approach is computationally expensive for large spatio-temporal data sets. Previous work has shown that instrument changes in VOD data due to multiple overlapping satellite data sources can induce biases in stability estimates \citep{smith2022b}; in short, changing signal-to-noise ratios between overlapping satellite data sets leave traces in both AC1 and variance that could be misinterpreted as stability changes. 

Our approach attempts to reconstruct the underlying attractor despite noise; DMD is well-suited to this task even under variable noise levels. There is no evidence of a trend, and hence a bias, in the eigenvalue magnitudes as instruments are added or removed (Supplemental Figure S1), despite clear biases in AC1 and variance. If noise levels are reasonable (\textless the magnitude of the signal), we still capture the CSD eigenvalue cleanly despite changing noise levels (Supplemental Figures S2, S4). We thus posit that VOD data  -- which has relatively high signal-to-noise ratios despite data mixing \citep{smith2022b} -- is suitable for analysis with our method. 

\subsection*{Greenland Melt Data}

We test our analysis on non-seasonal data that has been shown via other methods to exhibit signs of CSD \citep{boers2021}. This data is sourced from Trusel et al. \citep{trusel2018}, and is comprised of central west Greenland stack melt estimates from ice core records annually sampled since 1675 \citep{cappelen2014}. We do not perform any other pre-processing on the data -- we simply apply a rolling window ($t=100$ years) and estimate changes in the largest eigenvalue using both S-Map and DMD (Supplemental Figure S9). 

\section*{Data and Code Availability}

All data is publicly available. Vegetation data can be found at \url{https://doi.org/10.5281/zenodo.2575599} \citep{VOD}. Greenland melt data can be found via \url{https://www.nature.com/articles/s41586-018-0752-4\#Sec14} \citep{trusel2018}. ITS\_LIVE glacier velocity data was accessed via the public Python API \citep{gardner2018,gardner2025}. Synthetic data creation, glacier data access, and analysis scripts are publicly available on Zenodo: \textit{link to appear with publication}.

\section*{Acknowledgments}
T.S. acknowledges support from the DFG STRIVE project (SM 710/2-1) and the Universit\"{a}t Potsdam Remote Sensing Computational Cluster. N.B. acknowledges funding by the Volkswagen Foundation. This is ClimTip contribution \#X; the ClimTip project has received funding from the European Union's Horizon Europe research and innovation programme under grant agreement No. 101137601. This study received additional support from the European Space Agency Climate Change Initiative (ESA-CCI) Tipping Elements SIRENE project (contract no. 4000146954/24/I-LR). 

\section*{Author Contributions}
T.S. conceived and designed the study, processed the data, and performed the numerical analysis. T.S. wrote the manuscript with contributions from A.M., B.B., and N.B.

\section*{Competing Interests}
The authors declare no competing financial or non-financial interests.

\clearpage
\newpage
\singlespacing

\clearpage
\newpage

\pagenumbering{gobble}

\begin{centering}
	\clearpage\thispagestyle{empty}
	\textbf{\Large{Supplement to: Predicting Instabilities in Transient Landforms and Interconnected Ecosystems}} 
	\vspace{2cm}
	
	Taylor Smith$^{1}$, Andreas Morr$^{2,3}$, Bodo Bookhagen$^{1}$, Niklas Boers$^{2,3,4}$\\
	$^{1}$Institute of Geosciences, Universit\"{a}t Potsdam, Germany\\
	$^{2}$ Earth System Modelling, School of Engineering \& Design, Technical University of Munich, Germany  \\
	$^{3}$Potsdam Institute for Climate Impact Research, Germany\\
	$^{4}$Department of Mathematics and Global Systems Institute, University of Exeter, UK\\
	
\end{centering}

\vspace{16cm}

\noindent
Corresponding author: \\
Taylor Smith \\
Email: tasmith@uni-potsdam.de

\clearpage
\doublespacing

\setcounter{figure}{0}
\renewcommand{\figurename}{Supplementary Figure}
\renewcommand{\thefigure}{S\arabic{figure}}
\renewcommand{\thetable}{S\arabic{table}}

\clearpage
\newpage

\begin{figure*}[!h]
	\centering
	\includegraphics[width=\linewidth]{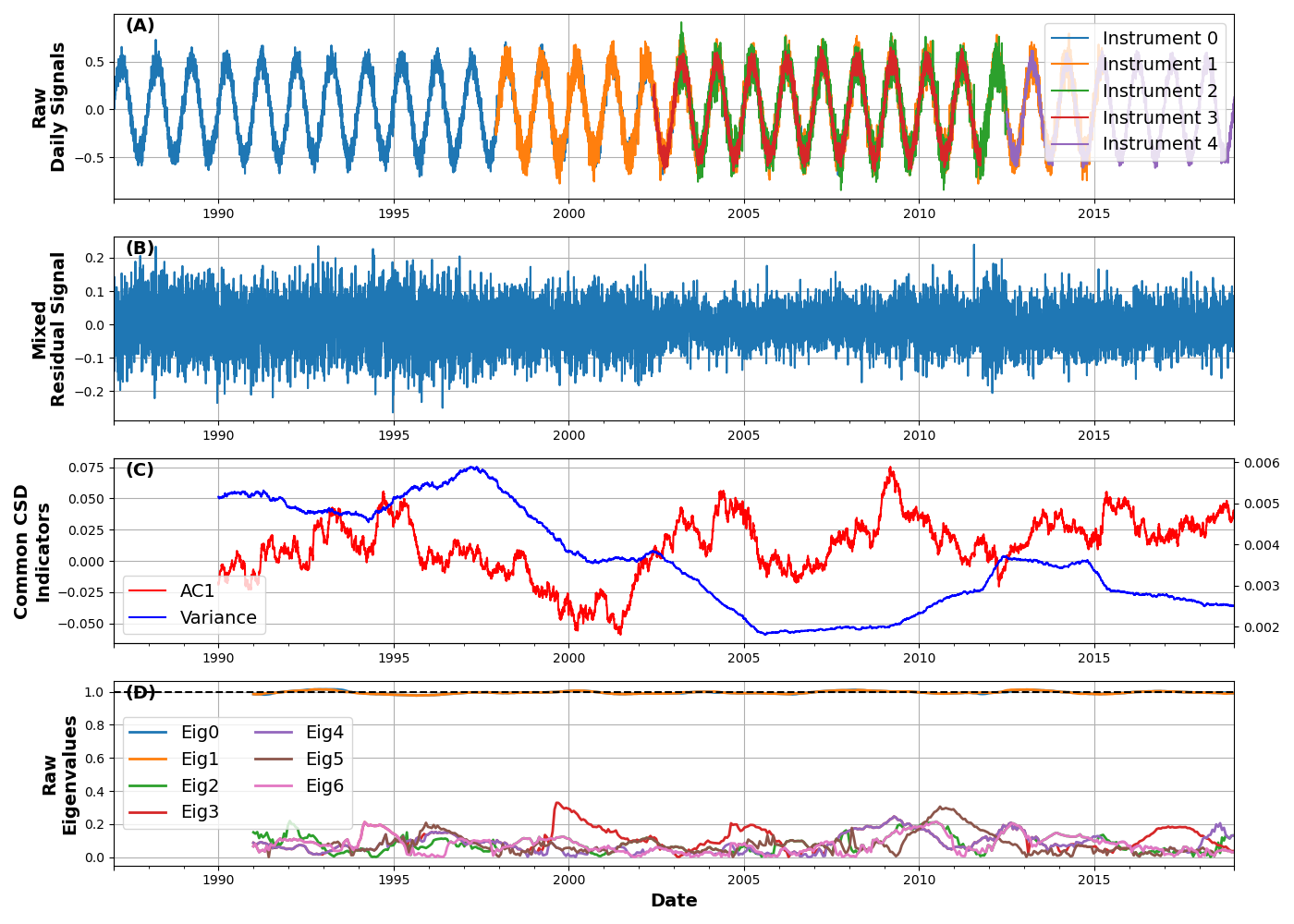}
	\caption{(A) Synthetic signals with varying noise levels and overlapping time periods. (B) Mixed daily average, deseasoned and detrended with STL. (C) AC1 and variance computed on the deseasoned residual, showing spurious changes due to noise level changes. (D) Eigenvalues computed on the raw daily signal without deseasoning or detrending, showing that the stable cycle is captured (strong eigenvalue near 1) and the noise changes are partitioned into low-energy eigenvalues. There are no spurious high-energy eigenvalues induced by noise-level changes.}
	\label{s1}
\end{figure*}

\begin{figure*}[!h]
	\centering
	\includegraphics[width=\linewidth]{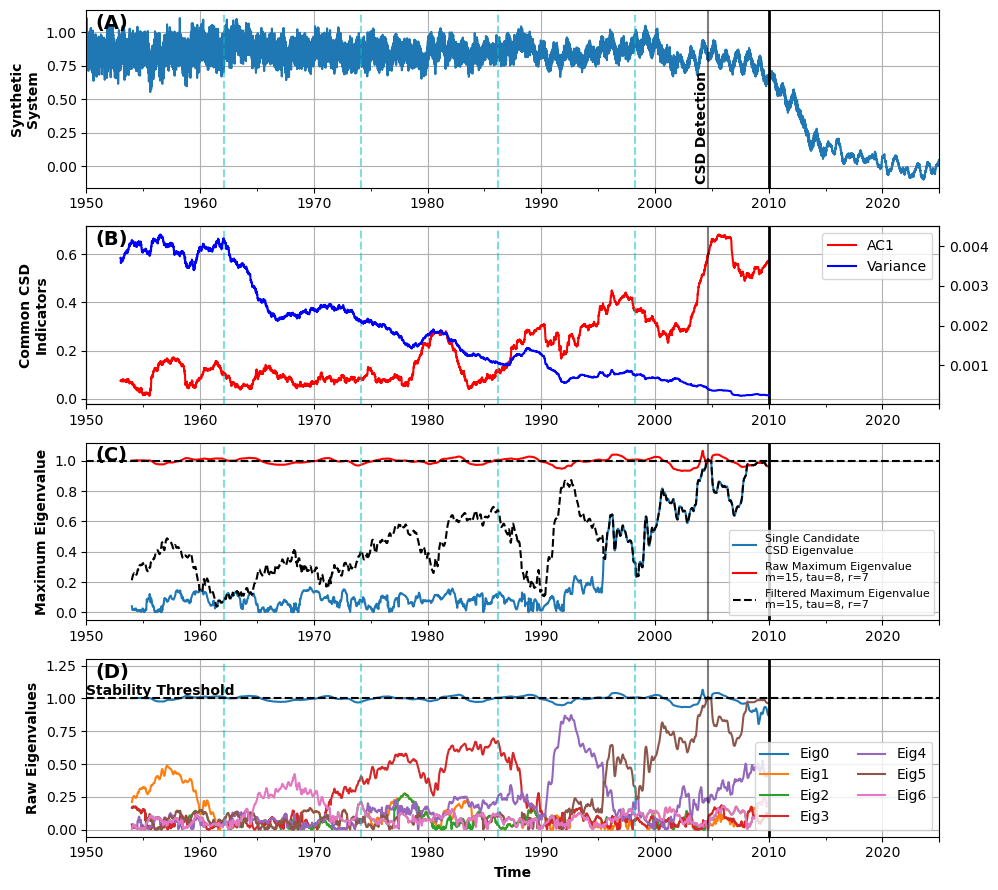}
	\caption{(A) Synthetic system with decreasing additive noise levels, marked at cyan dashed vertical lines. Vertical black line marks instability detection from eigenvalue tracking. (B) Changes in AC1 and variance due to noise level shifts in deseasoned and detrended residual. (C) Maximum and filtered maximum eigenvalue, showing expected CSD signal without bias. (D) All eigenvalues computed on the raw, noise-varying system.}
	\label{s2}
\end{figure*}

\begin{figure*}[!h]
	\centering
	\includegraphics[width=\linewidth]{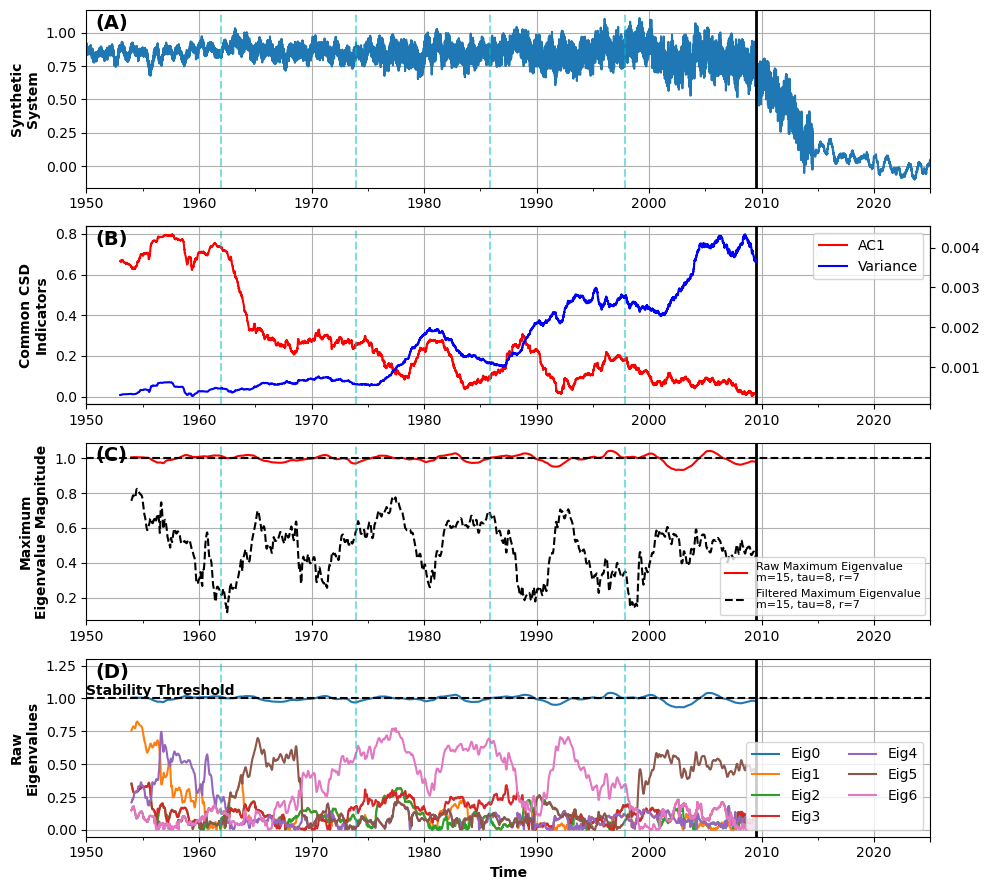}
	\caption{(A) Synthetic system with increasing additive noise levels, marked at cyan dashed vertical lines. (B) Changes in AC1 and variance due to noise level shifts in deseasoned and detrended residual. (C) Maximum and filtered maximum eigenvalue, showing a missed CSD signal due to high noise around the transition. (D) All eigenvalues computed on the raw, noise-varying system.}
	\label{s3}
\end{figure*}

\begin{figure*}[!h]
	\centering
	\includegraphics[width=\linewidth]{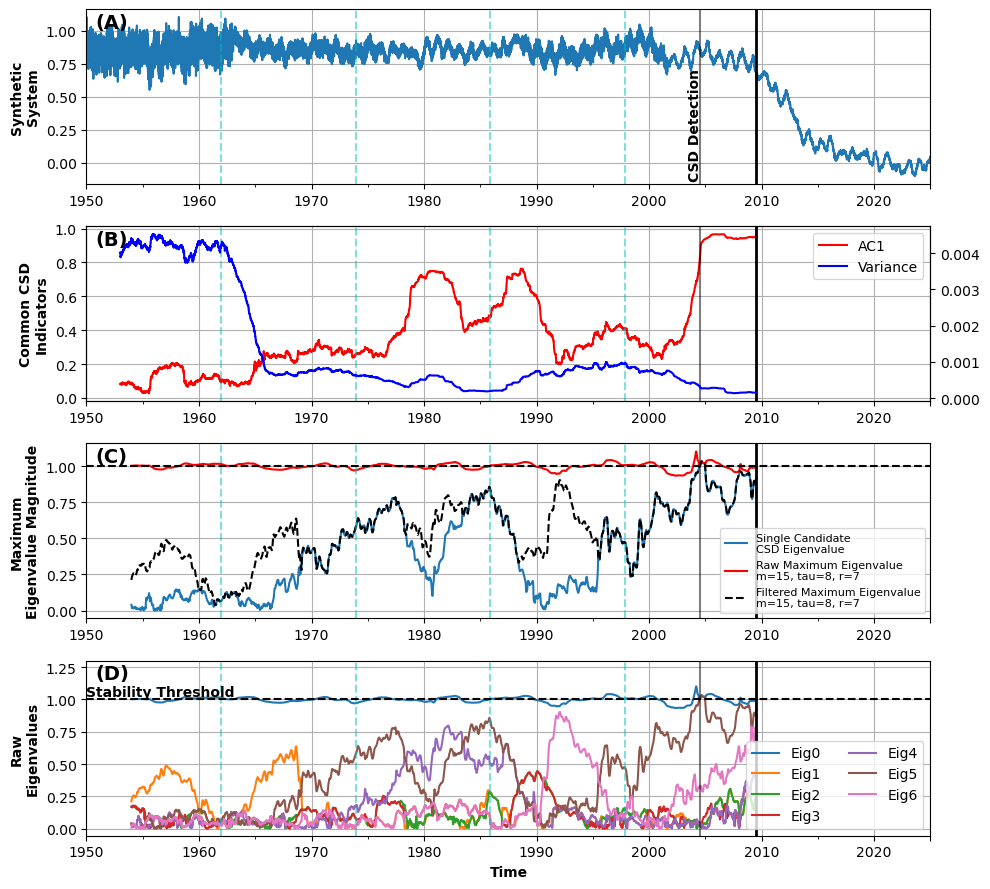}
	\caption{(A) Synthetic system with shuffled additive noise levels, marked at cyan dashed vertical lines. Vertical black line marks instability detection from eigenvalue tracking. (B) Changes in AC1 and variance due to noise level shifts in deseasoned and detrended residual. (C) Maximum and filtered maximum eigenvalue, showing expected CSD signal without bias. (D) All eigenvalues computed on the raw, noise-varying system.}
	\label{s4}
\end{figure*}

\begin{figure*}[!h]
	\centering
	\includegraphics[width=\linewidth]{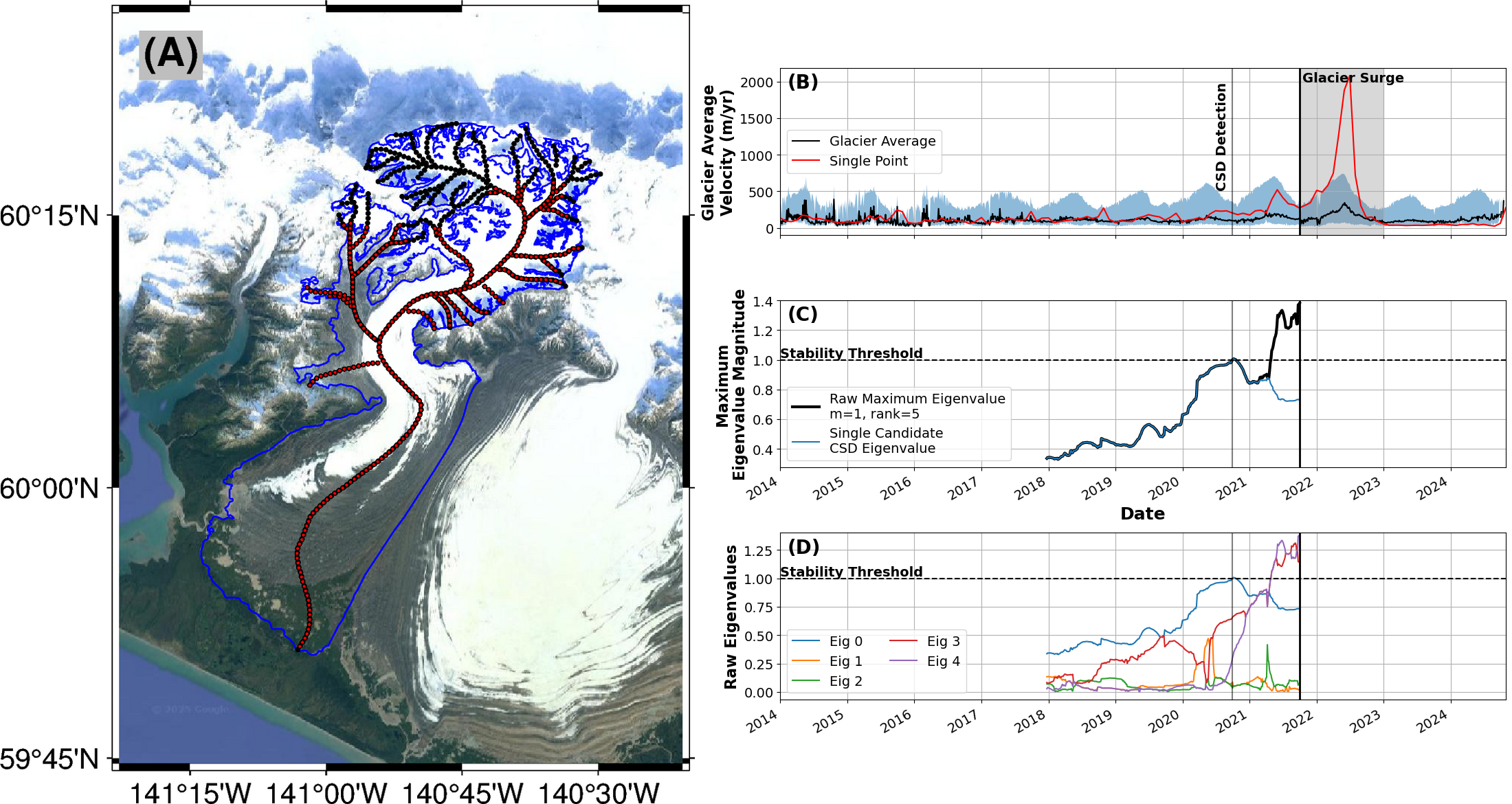}
	\caption{(A) Map of target glacier (RGI2000-v7.0-G-01-13271, [-140.847,60.089]) with outline (blue) and sample points ($n=337$) along the glacier centerline (red, low-elevation points, black, all centerline points). Background from Google Earth. (B) 4-day ($t=1004$ time steps) glacier-average velocity (black) plus 25th-75th percentile range (shaded blue), alongside single strongly surging point (red). Shaded area shows surge period based on velocity magnitude. Vertical black line marks instability detection from eigenvalue tracking. (C) Maximum eigenvalue using the set of velocities over all low-elevation centerline points, showing a trend towards instability far before the actual surge despite lack of lag embedding ($m$=1, $r$=5). (D) All computed eigenvalues.}
	\label{s5}
\end{figure*}

\begin{figure*}[!h]
	\centering
	\includegraphics[width=\linewidth]{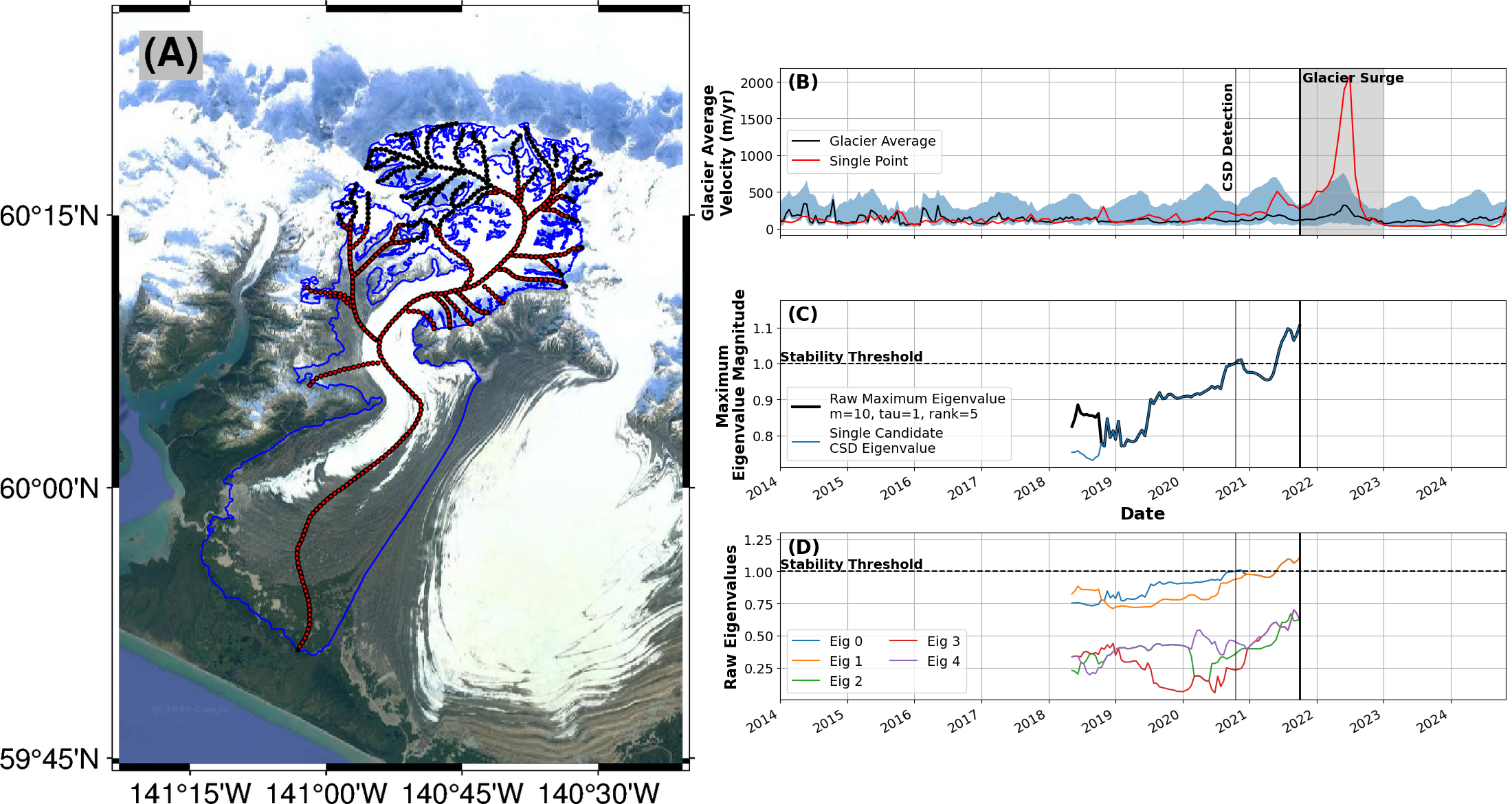}
	\caption{(A) Map of target glacier (RGI2000-v7.0-G-01-13271, [-140.847,60.089]) with outline (blue) and sample points ($n=337$) along the glacier centerline (red, low-elevation points, black, all centerline points). Background from Google Earth. (B) 16-day ($t=253$ time steps) glacier-average velocity (black) plus 25th-75th percentile range (shaded blue), alongside single strongly surging point (red). Shaded area shows surge period based on velocity magnitude. Vertical black line marks instability detection from eigenvalue tracking. (C) Maximum eigenvalue using the set of velocities over all low-elevation centerline points, showing a trend towards instability far before the actual surge. (D) All computed eigenvalues ($m$=10, $\tau$=1, $r$=5).}
	\label{s6}
\end{figure*}

\begin{figure*}[!h]
	\centering
	\includegraphics[width=\linewidth]{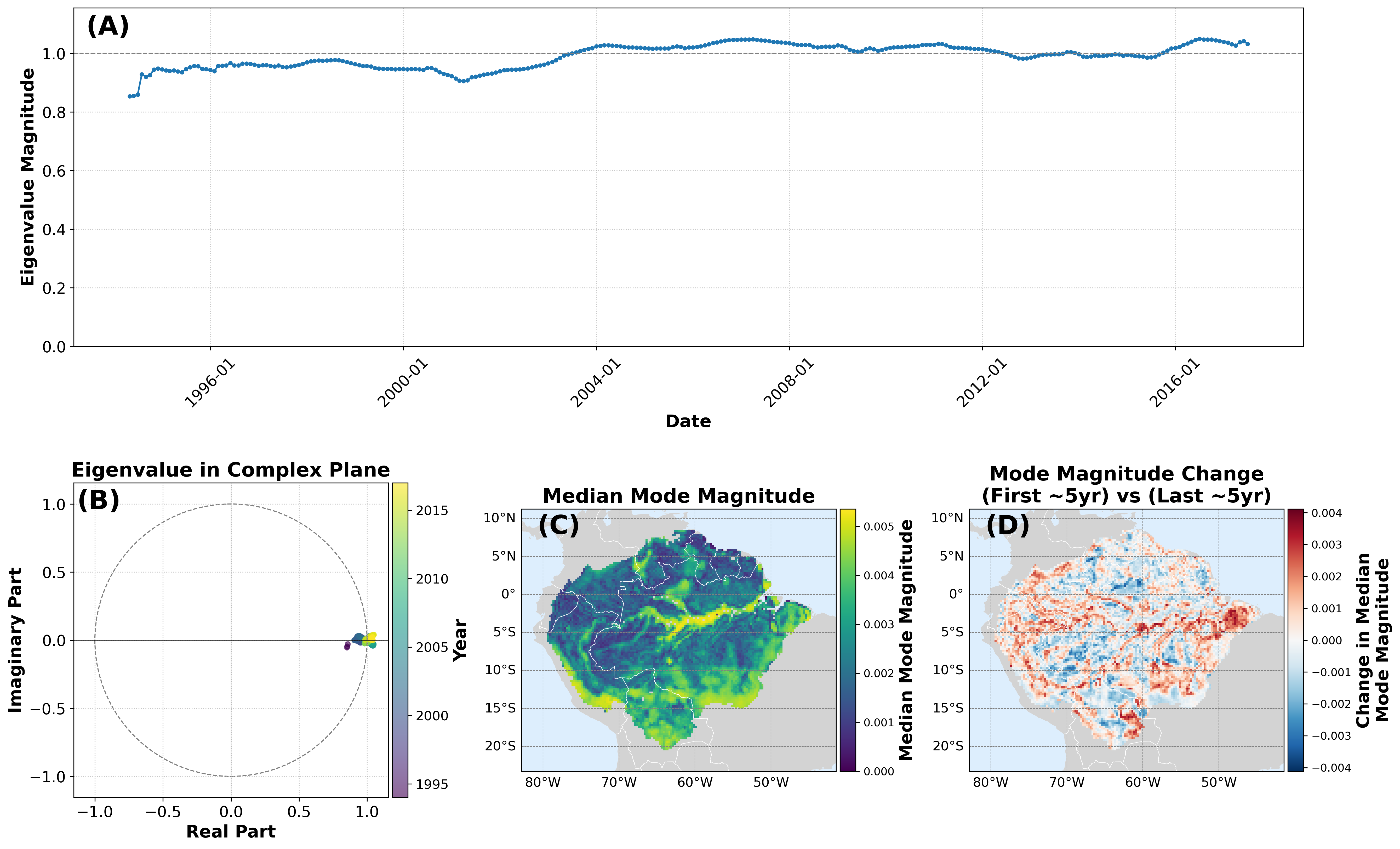}
	\caption{(A) Single periodic eigenvalue of interest. (B) Eigenvalue track in the complex plane, showing a constant location along the unit circle. (C) Spatial map of the median spatial mode associated with the eigenvalue track of interest. (D) Change in the spatial mode map between the first and last five years of data.}
	\label{s7}
\end{figure*}

\begin{figure*}[!h]
	\centering
	\includegraphics[width=\linewidth]{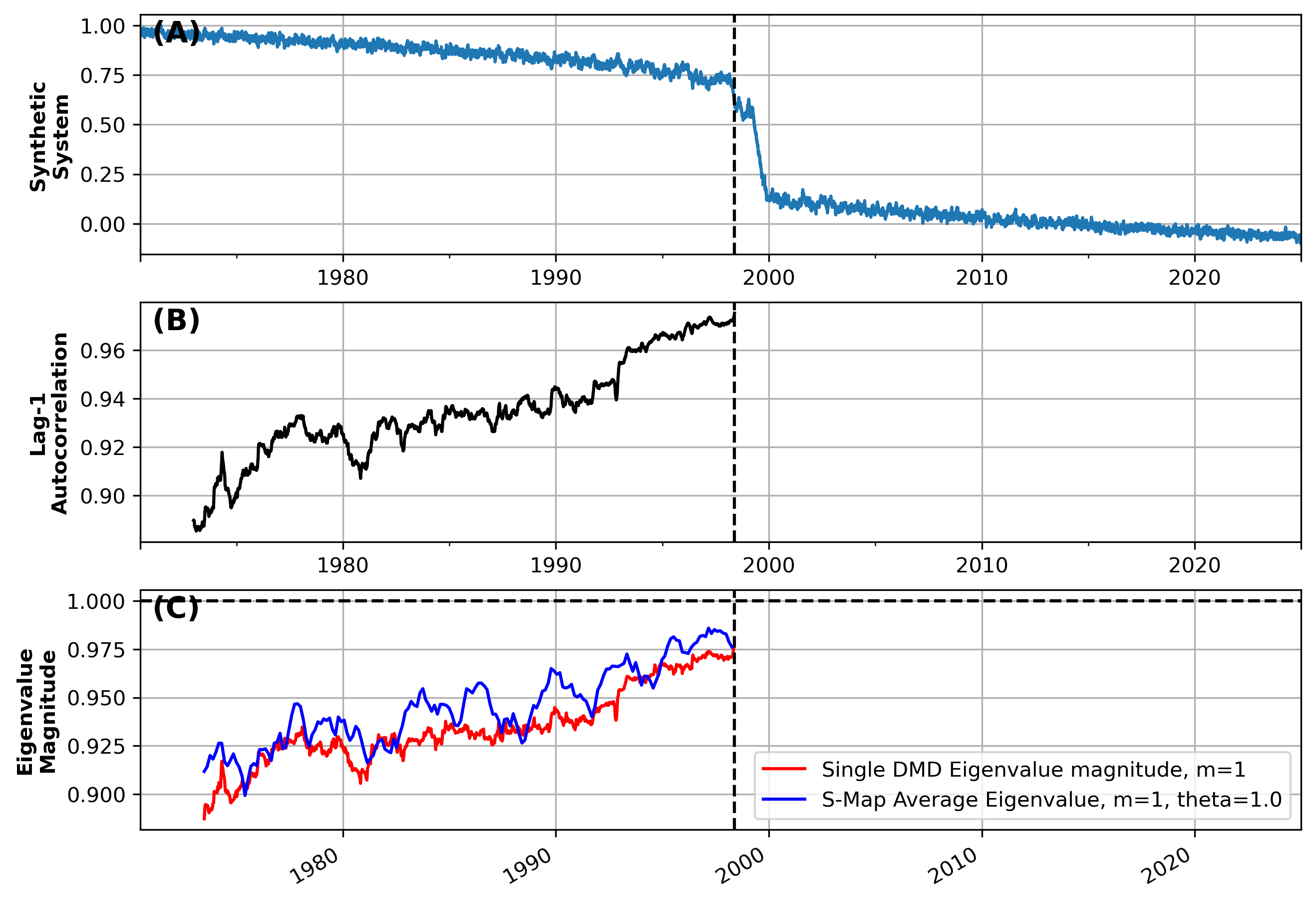}
	\caption{(A) Toy model showing pitchfork bifurcation. (B) Lag-1 autocorrelation over a set of moving windows and (C) DMD and S-Map dominant eigenvalue estimates for the same set of moving windows without lag embedding.}
	\label{s8}
\end{figure*}

\begin{figure*}[!h]
	\centering
	\includegraphics[width=\linewidth]{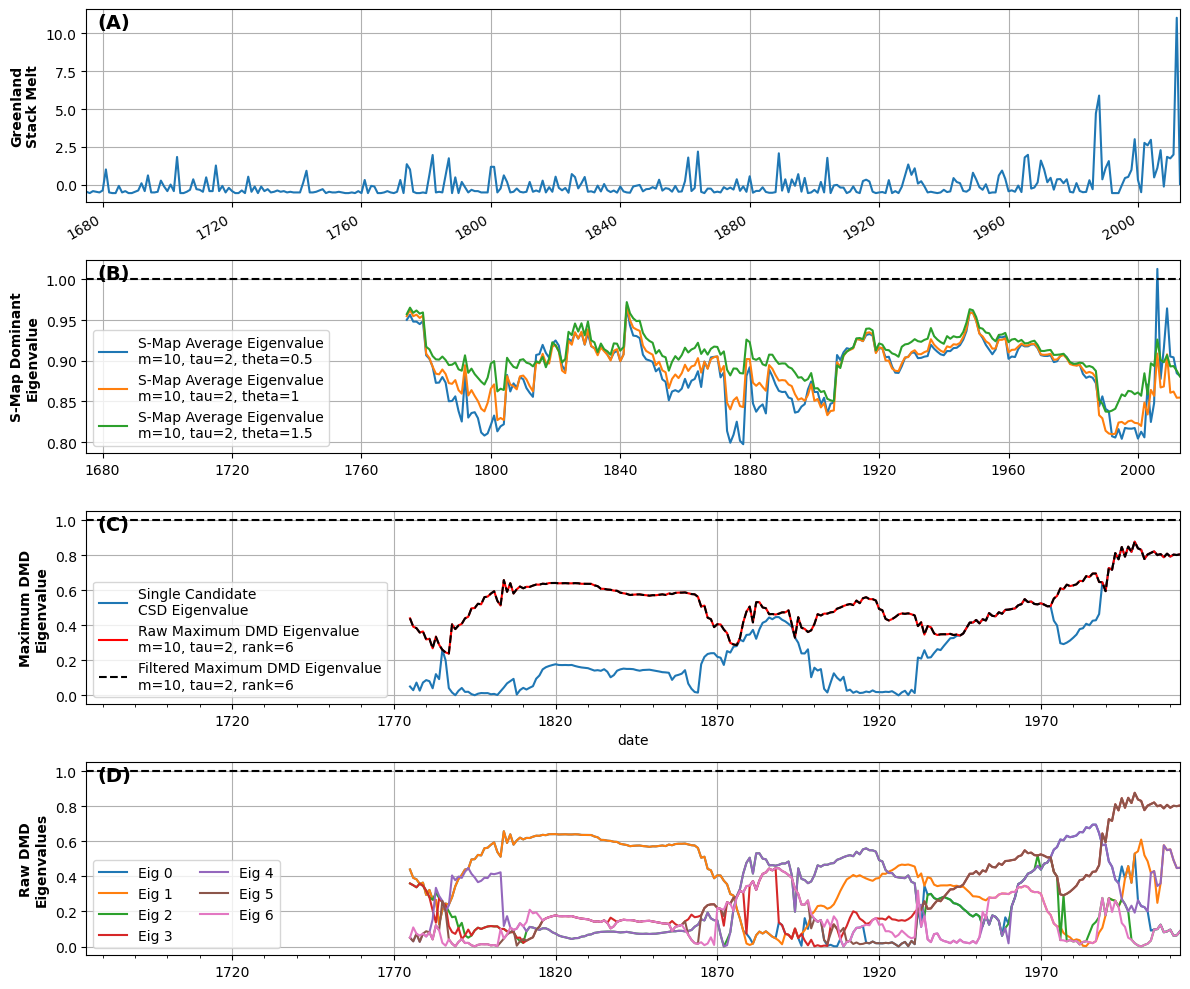}
	\caption{(A) Greenland Stack Melt. (B) S-Map eigenvalues for three different theta parameters. (C) Filtered maximum DMD eigenvalue and (D) all DMD eigenvalues.}
	\label{s9}
\end{figure*}

\begin{figure*}[!h]
	\centering
	\includegraphics[width=\linewidth]{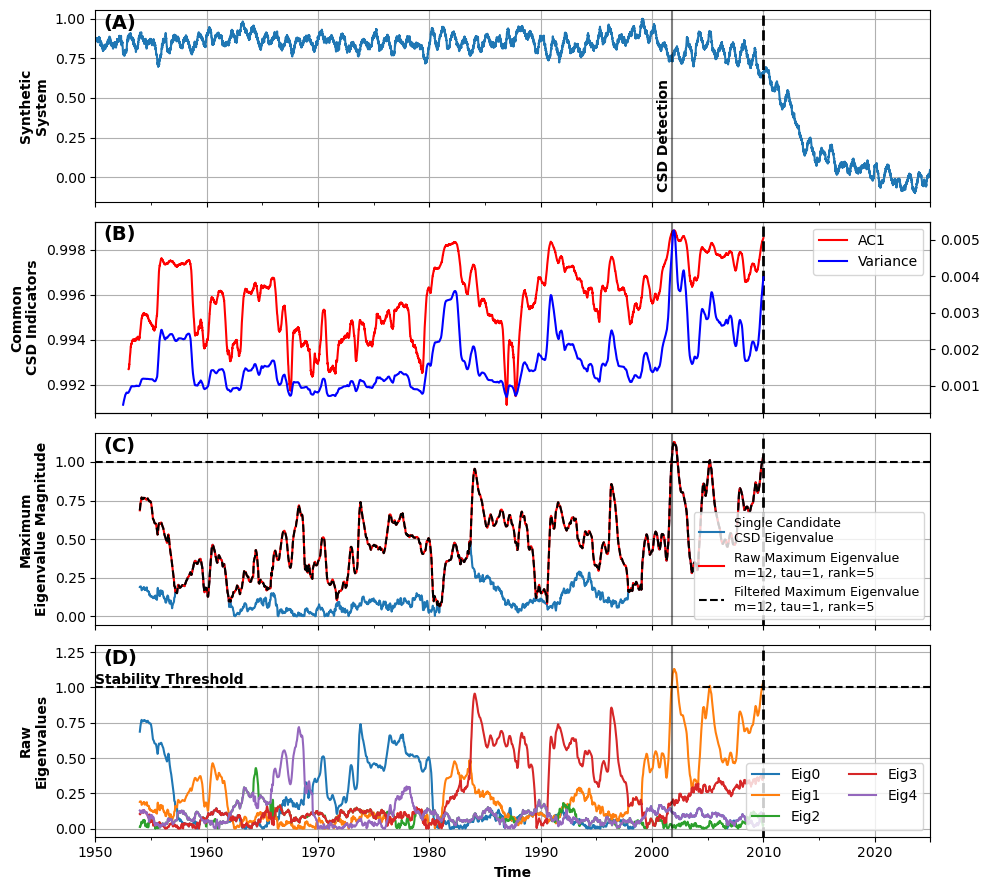}
	\caption{(A) Simple model with seasonality moving towards a state transition. Vertical black line marks instability detection from eigenvalue tracking. (B) Lag-one autocorrelation and variance computed directly on the time series without deseasoning. (C) Filtered and dominant eigenvalues with mean centering, showing that a dominant eigenvalue approaches and crosses 1 before the critical transition. There is no longer a stable single periodic eigenvalue close to 1. (D) The set of all eigenvalues.}
	\label{s10}
\end{figure*}

\begin{figure*}[!h]
	\centering
	\includegraphics[width=0.75\linewidth]{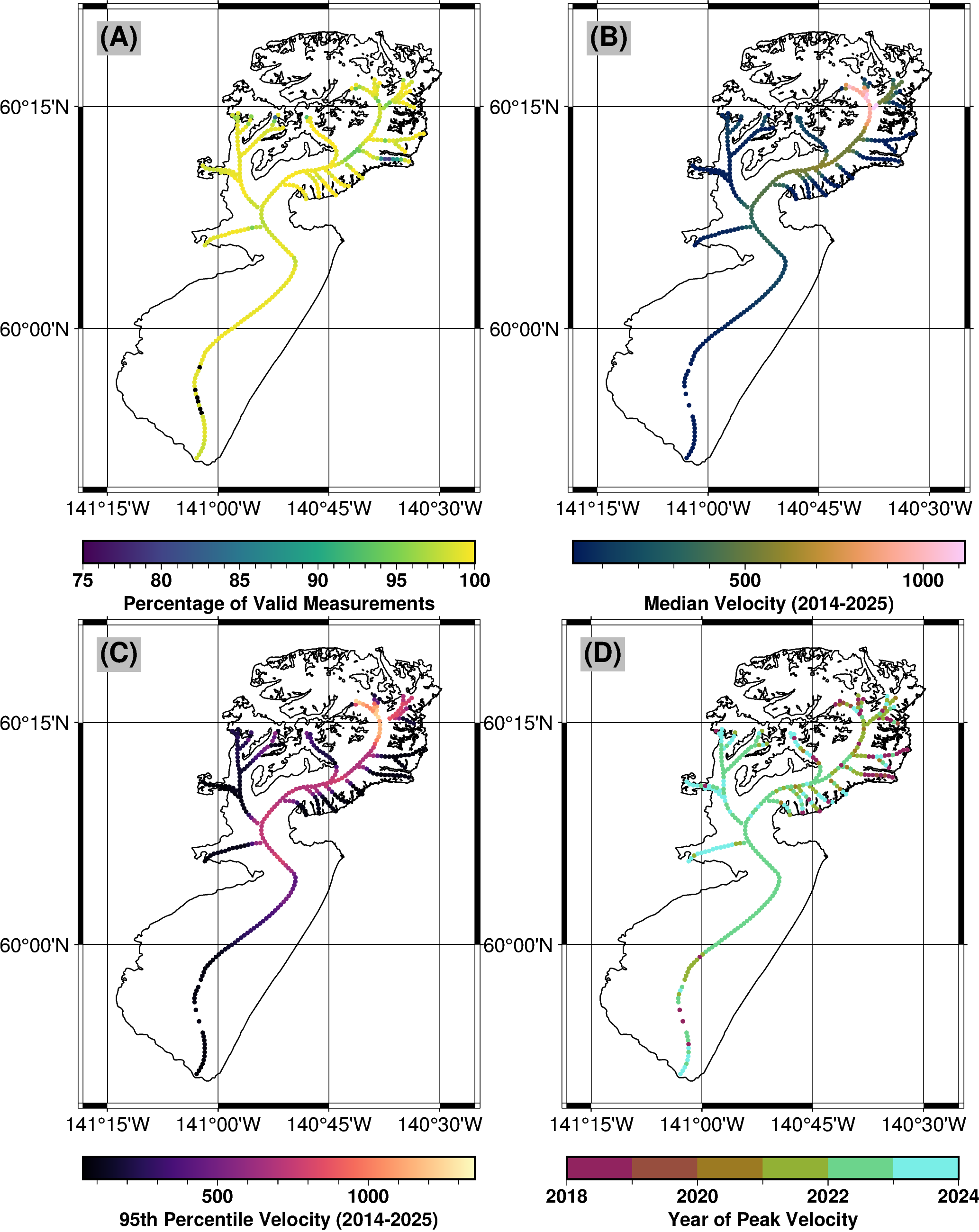}
	\caption{Spatial patterns of glacier velocity data for RGI2000-v7.0-G-01-13271. (A) Percentage of available measurements per point. (B) Median velocity over the whole data period. (C) 95th percentile velocity. (D) Year of maximum glacier velocity.}
	\label{s11}
\end{figure*}

\end{document}